\documentclass[12pt,a4paper]{article}
\usepackage{amsmath,amssymb,amsthm}
\usepackage{hyperref}
\usepackage{geometry}
\title{Spin--Induced Geometry: Emergence of Metric and\\
Torsional Sectors from Spinor Source}

\author{Elisa Varani\\
\small Universit\`a Cattolica del Sacro Cuore, Milan, Italy\\
\small \texttt{elisa.varani@unicatt.it}}

\date{}

\begin{document}
\maketitle

\begin{abstract}
We present a geometric framework in which both metric and torsional degrees of freedom
emerge dynamically from spinor currents, without being postulated as fundamental properties
of the affine connection. The fundamental dynamical variable is a rank-three field carrying
local Lorentz indices, governed by a massive Klein--Gordon equation sourced by fermionic spin
currents. The microscopic Clifford source directly excites the axial torsional sector. A
symmetric metric response may emerge at the effective level through coherent axial
spin-current domains, whose collective order parameter induces an effective geometric
response. The symmetric and antisymmetric sectors define, respectively, an effective
spin-induced metric and the torsional degrees of freedom. Both sectors are massive and
Yukawa-suppressed, ensuring decoupling from long-range gravitational dynamics. Unlike
Einstein--Cartan theory, torsion here is propagating rather than algebraically constrained.
A key consequence is that spinless test particles follow geodesics of the effective metric and
are therefore indirectly sensitive to spin currents through the emergent geometric structure
--- a mechanism absent in both standard General Relativity and Einstein--Cartan theory.
The spinorial structure of the source is analyzed across three regimes: general Dirac, Weyl,
and Majorana fermions, each giving rise to a distinct geometric phase. In the Majorana limit, and in the absence of coherent axial ordering,
the geometry becomes purely axial-torsional, admitting topologically non-trivial configurations
such as vortices and Skyrmion-like structures, which emerge dynamically from the spinorial
source~\cite{varani2025ricci}.
\end{abstract}


\section{Introduction}

\subsection{Spin-Induced Geometry}

We examine the compatibility of a broader spin--driven geometric framework with General
Relativity and Einstein-Cartan theory~\cite{hehl1976,trautman2006}. In this perspective,
geodesic and autoparallel trajectories do not represent independent postulates, but rather
distinct manifestations of the same underlying spin--generated geometry. The symmetric sector
governs the effective metric and the associated geodesic structure, while the antisymmetric
sector encodes torsion, affine effects, and topological features.

In previous works by the author~\cite{varani2022,varani2025yukawa}, it was demonstrated that
fermionic spin currents can dynamically generate torsion-like geometric effects. Within a
Riemannian background endowed with a Levi--Civita connection, the tetradic Hamiltonian
formulation yields a canonical fermionic rotational current, identified with the momentum
conjugate to the tetrad field. This current represents the dynamical response of fermionic
spin to local Lorentz rotations of the tetrad field,
\begin{equation}
s_{cab} = -\frac{i}{4}\,\bar\psi\,\{\gamma_c,\sigma_{ab}\}\,\psi,
\label{eq:scab}
\end{equation}
which possesses the same algebraic structure as the spin tensor appearing in
Einstein--Cartan theory. This result motivates the possibility that torsion-like effects may
arise dynamically from spinor currents, without being postulated as a fundamental geometric
property of the affine connection.

Chiral fermions further generate a dynamical structure of the form
\begin{equation}
\tilde\phi = h^{\mu c|b}\{\gamma_\mu, \sigma_{cb}\},
\end{equation}
where the tensor $h^{\mu c|b}$ behaves as a dynamical field. This structure arises explicitly
from the spin-connection terms appearing in the linearized Dirac Lagrangian, where derivatives
of the field $h_{\mu c|b} = \partial_b h_{\mu c}$ couple to fermionic spin through the
anticommutator $\{\gamma^\mu, \sigma_{cb}\}$~\cite{varani2025yukawa}.

When this contribution enters the affine geometric structure of spacetime, the resulting
geometry departs from the purely Riemannian regime: the affine connection acquires torsional
and non-metric components, and physical trajectories are described by autoparallels of the
full affine connection, which no longer coincide with metric geodesics and become sensitive
to the internal spin structure of matter~\cite{shapiro2002}. In this respect, the notion of
physical trajectories is formally analogous to that of Einstein--Cartan theory and more
generally to metric-affine theories of gravity~\cite{hehl1995}, although the origin and
dynamical nature of torsion are fundamentally different.

The resulting framework admits propagating torsional modes and nontrivial topological
configurations, such as vortices and skyrmion-like structures, which are absent in standard
General Relativity. By projecting the geometric structure onto its symmetric metric sector,
metric compatibility and a Levi--Civita connection are recovered.

\section{Materials and Methods}

\subsection{Dynamical Torsion and Geodesic Deviation}

This section defines the theoretical and geometric foundations required to analyze
gravitational dynamics in the presence of spinor-induced perturbations. We study the
consequences of decomposing the perturbation field $h_{\mu\nu}$ into its symmetric and
antisymmetric components. The massive nature of the perturbation field leads to
Yukawa--type suppression at the microscopic level, confining local non--Riemannian effects
to short distances. At the same time, this does not preclude the emergence of effective
macroscopic geometric features due to collective behavior.

\subsection{Spin Current Structure and World-Index Projection}
\label{sec:spincurrent}

In this subsection, we provide an explicit construction of the spacetime metric and torsional
contributions induced by the spinorial source. At the fundamental level, the geometric
perturbations generated by spin degrees of freedom are formulated in terms of fields carrying
local Lorentz indices. This fundamental field, $h_{\mu c|b}$, represents a unified geometric
structure that encompasses both metric and torsional degrees of freedom. The dynamics of this
sector is governed by a massive Klein-Gordon-type equation sourced by spin
currents~\cite{varani2025yukawa},
\begin{equation}
\Box\,h_{\mu c|b} - 8m^2\,h_{\mu c|b} = S_{\mu cb}(x),
\label{eq:KG_lorentz}
\end{equation}
with source term
\begin{equation}
S_{\mu cb}(x) = \frac{2i\,\bar\psi\,\{\gamma_\mu,\sigma_{cb}\}\,\psi}
{\{\gamma_\mu,\sigma_{cb}\}^2},
\label{eq:source_lorentz}
\end{equation}
where the denominator is understood as the fully contracted Clifford scalar
\begin{equation}
\{\gamma_\mu,\sigma_{cb}\}^2 \equiv
\{\gamma_\mu,\sigma_{cb}\}\{\gamma^\mu,\sigma^{cb}\} = C\,\mathbf{1},
\end{equation}
which is proportional to the identity in spinor space and therefore reduces to a real positive
constant $C > 0$.

The specific symmetry properties of the gravitational and torsional sectors emerge
dynamically from the index structure of the fundamental field $h_{\mu c|b}$ and its coupling
to the spinorial source.

To relate this formulation to spacetime observables, we work within the weak--field
approximation, in which the tetrad field $e^a{}_\mu$ is expanded around the Minkowski
background as
\begin{equation}
e^a{}_\mu = \delta^a{}_\mu + \tfrac{1}{2}\,h^a{}_\mu,
\end{equation}
with $|h^a{}_\mu| \ll 1$. To first order in perturbations, the projection from local Lorentz
indices to spacetime indices becomes purely algebraic,
\begin{equation}
e^a{}_\mu \simeq \delta^a{}_\mu.
\end{equation}
The passage from Lorentz to spacetime indices is achieved by contraction with the tetrad
fields,
\begin{equation}
h_{\mu\nu|\lambda} \equiv e^c{}_\nu\,e^b{}_\lambda\,h_{\mu c|b}
\simeq \delta^c{}_\nu\,\delta^b{}_\lambda\,h_{\mu c|b},
\end{equation}
which reduces to a purely algebraic identification in the weak--field limit.

We then introduce the spacetime field
\begin{equation}
h_{\mu\nu|\lambda}(x) \equiv \partial_\lambda h_{\mu\nu}(x),
\end{equation}
which satisfies the Klein--Gordon--type equation with spinorial source, written in world
indices,
\begin{equation}
\bigl(\Box - 8m^2\bigr)\,h_{\mu\nu|\lambda}(x) = S_{\mu\nu\lambda}(x),
\label{eq:KG_world}
\end{equation}
where
\begin{equation}
S_{\mu\nu\lambda}(x) =
\frac{2i\,\bar\psi(x)\,\{\gamma_\mu,\sigma_{\nu\lambda}\}\,\psi(x)}
{\{\gamma_\mu,\sigma_{\nu\lambda}\}^2}.
\label{eq:source_world}
\end{equation}
Since
\begin{equation}
\sigma_{\nu\lambda} = -\sigma_{\lambda\nu},
\end{equation}
it follows immediately that
\begin{equation}
h_{\mu\nu|\lambda} = -h_{\mu\lambda|\nu},
\end{equation}
so that the fundamental field is antisymmetric in its last two indices.

The non--homogeneous solution can be written in terms of the Green function as
\begin{equation}
h_{\mu\nu|\lambda}(x) = \int d^4x'\,G(x,x')\,S_{\mu\nu\lambda}(x').
\end{equation}
A rank--2 spacetime tensor is obtained by reconstructing the field through integration over
the derivative index,
\begin{equation}
h_{\mu\nu}(x) \equiv \int dx'^\lambda\,h_{\mu\nu|\lambda}(x'),
\label{eq:hmunu_reconstruct}
\end{equation}
where the integration reconstructs $h_{\mu\nu}$ from its derivative representation
$h_{\mu\nu|\lambda} = \partial_\lambda h_{\mu\nu}$.

Using the four-dimensional Clifford algebra, the anticommutator entering the spinorial source
selects the axial channel,
\begin{equation}
\{\gamma_\mu,\sigma_{\nu\lambda}\} = C_\epsilon\,\varepsilon_{\mu\nu\lambda\rho}\,
\gamma^\rho\gamma^5,
\label{eq:anticomm_axial}
\end{equation}
where $C_\epsilon$ depends on the conventions adopted for $\gamma^5$,
$\varepsilon_{\mu\nu\lambda\rho}$, and $\sigma_{\mu\nu}$. Therefore the microscopic
spinorial source is
\begin{equation}
S^{(A)}_{\mu\nu\lambda} = \kappa_A\,\varepsilon_{\mu\nu\lambda\rho}\,J_A^\rho,
\qquad J_A^\rho = \bar\psi\,\gamma^\rho\gamma^5\,\psi.
\label{eq:source_axial}
\end{equation}
This source is totally antisymmetric in its three tensorial indices and therefore directly
feeds the axial torsional sector. The symmetric metric sector is instead associated with the effective trace-vector response
$S^{\rm eff}_{\mu\nu\lambda}$ generated by coherent axial-current domains, as defined below.

In order to describe the metric response generated by collective spinorial configurations,
we introduce the \emph{coherent axial order vector}
\begin{equation}
U^\mu(x) = \langle J_A^\mu(x)\rangle_{\rm coh}.
\label{eq:order_vector}
\end{equation}
For randomly oriented microscopic spin currents, $U^\mu = 0$. In a coherent domain, however,
$U^\mu \neq 0$, and the axial spin current defines a macroscopic vectorial direction. At the level of the affine response, the torsion tensor admits the standard
irreducible Lorentz decomposition into axial, trace-vector, and purely tensorial components~\cite{hehl1976,shapiro2002}. While the Clifford anticommutator fixes the axial component, a
coherent axial order parameter may polarize the trace-vector channel of the
effective torsional response.
The trace-vector component has the standard form
\begin{equation}
T^{(V)}_{\mu\nu\lambda}
=
\frac{1}{3}
\left(g_{\mu\nu}V_\lambda-g_{\mu\lambda}V_\nu\right),
\qquad
V_\lambda=T^\rho{}_{\rho\lambda}.
\end{equation}
At the effective level, the coherent axial order parameter may induce a
trace-vector response, which we parameterize as
\[
V_\lambda=\alpha U_\lambda ,
\]
where \(\alpha\) is a phenomenological coupling constant.
After absorbing the numerical factor and \(\alpha\) into an effective coupling \(\chi\), the effective response is
\begin{equation}
S^{\rm eff}_{\mu\nu\lambda}
= \chi\bigl(g_{\mu\nu}U_\lambda - g_{\mu\lambda}U_\nu\bigr),
\label{eq:Seff}
\end{equation}
where $\chi$ is an effective coupling constant. This object represents the macroscopic trace-vector response induced by coherent axial
spin currents. The two-level structure of the model is therefore:
\begin{equation}
S^{(A)}_{\mu\nu\lambda} \;\Rightarrow\; h_{[\mu\nu]},
\qquad
S^{\rm eff}_{\mu\nu\lambda}(U) \;\Rightarrow\; h_{(\mu\nu)}.
\label{eq:two_level}
\end{equation}

The projection onto the metric sector is then achieved by selecting the symmetric component,
\begin{equation}
h_{(\mu\nu)} \equiv \tfrac{1}{2}(h_{\mu\nu}+h_{\nu\mu}),
\label{eq:symm}
\end{equation}
which defines the effective spinor--induced contribution to the spacetime metric. The
antisymmetric component $h_{[\mu\nu]}$ does not define metric degrees of freedom and remains
associated with the axial torsional sector.

It is important to note that the fundamental dynamical variable carries three indices,
$h_{\mu\nu|\lambda} = \partial_\lambda h_{\mu\nu}$. The symmetric spacetime metric component
therefore emerges only after the reconstruction of the field; we consider a detailed analysis
in the next section.

\subsection{Symmetric and Antisymmetric Decomposition}
\label{sec:decomp}

Any rank--2 tensor admits a unique decomposition into symmetric and antisymmetric components,
\begin{equation}
h_{\mu\nu} = h_{(\mu\nu)} + h_{[\mu\nu]},
\end{equation}
where
\begin{align}
h_{(\mu\nu)} &\equiv \tfrac{1}{2}(h_{\mu\nu}+h_{\nu\mu}),\\
h_{[\mu\nu]} &\equiv \tfrac{1}{2}(h_{\mu\nu}-h_{\nu\mu}).
\end{align}

Taking into account the definition~\eqref{eq:hmunu_reconstruct}, the symmetric component
can be written as
\begin{equation}
h^{\rm eff}_{(\mu\nu)}(x)
=
\frac{1}{2}
\int d^4x'\,G(x,x')
\bigl[
J^{\rm eff}_{\mu\nu}(x')
+
J^{\rm eff}_{\nu\mu}(x')
\bigr],
\label{eq:symm_component}
\end{equation}
where the effective rank--two source is defined by
\begin{equation}
J^{\rm eff}_{\mu\nu}(x)
\equiv
\int dx'^\lambda\,S^{\rm eff}_{\mu\nu\lambda}(x').
\label{eq:Jeff}
\end{equation}

To generate the symmetric sector, one must evaluate the collective coherent
effect as a trace-vector response source, whereas the microscopic axial Clifford
source contributes directly to the antisymmetric torsional sector. We therefore write,
as in Eq.~\eqref{eq:Seff},
\begin{equation}
S^{\rm eff}_{\mu\nu\lambda}
=
\chi
\left(
g_{\mu\nu}U_\lambda
-
g_{\mu\lambda}U_\nu
\right),
\label{eq:Seff_decomp}
\end{equation}
where
\begin{equation}
U_\lambda
=
\left\langle J^A_\lambda \right\rangle_{\rm coh}
=
\left\langle
\bar\psi\gamma_\lambda\gamma^5\psi
\right\rangle_{\rm coh}
\end{equation}
is the coherent axial spin-current order parameter.

Substituting the coherent order-parameter projection of the spinorial source,
one obtains
\begin{equation}
h^{\rm eff}_{(\mu\nu)}(x)
=
2i
\int d^4x'\,G(x,x')
\int dx'^\lambda
\Bigl[
\left\langle
\bar\psi(x')\{\gamma_\mu,\sigma_{\nu\lambda}\}\psi(x')
\right\rangle_{\rm coh,eff}
+
\left\langle
\bar\psi(x')\{\gamma_\nu,\sigma_{\mu\lambda}\}\psi(x')
\right\rangle_{\rm coh,eff}
\Bigr].
\label{eq:heff_spinorial_projection}
\end{equation}
Here \(\langle\cdots\rangle_{\rm coh,eff}\) denotes the coherent effective
projection of the microscopic spinorial source onto the trace-vector response
induced by the axial order parameter.

Similarly, the antisymmetric component associated with the microscopic axial
Clifford source is given by
\begin{equation}
h^{(A)}_{[\mu\nu]}(x)
=
\frac{1}{2}
\int d^4x'\,G(x,x')
\bigl[
J^{(A)}_{\mu\nu}(x')
-
J^{(A)}_{\nu\mu}(x')
\bigr],
\end{equation}
where
\begin{equation}
J^{(A)}_{\mu\nu}(x)
\equiv
\int dx'^\lambda\,S^{(A)}_{\mu\nu\lambda}(x').
\end{equation}
Upon substitution of the microscopic Clifford source, this contribution reads
\begin{equation}
h^{(A)}_{[\mu\nu]}(x)
=
2i
\int d^4x'\,G(x,x')
\int dx'^\lambda
\Bigl[
\bar\psi(x')\{\gamma_\mu,\sigma_{\nu\lambda}\}\psi(x')
-
\bar\psi(x')\{\gamma_\nu,\sigma_{\mu\lambda}\}\psi(x')
\Bigr].
\end{equation}

For the anticommutator source considered here, the direct microscopic contribution is axial,
\begin{equation}
S^{(A)}_{\mu\nu\lambda}
=
\kappa_A
\varepsilon_{\mu\nu\lambda\rho}
J_A^\rho,
\qquad
J_A^\rho
=
\bar\psi\gamma^\rho\gamma^5\psi .
\end{equation}
Thus, the antisymmetric component is naturally associated with axial torsion, while the
symmetric component is associated with the coherent axial-current response defined in
Eq.~\eqref{eq:Seff}.

The decomposition into \(h_{(\mu\nu)}\) and \(h_{[\mu\nu]}\) separates the reconstructed field
into an effective metric sector and a non-Riemannian torsional sector. The symmetric metric
perturbation is sourced by the coherent axial-current response, whereas the microscopic
anticommutator source directly generates the axial torsional sector.

While \(h^{\rm eff}_{(\mu\nu)}\) provides the gravitational potential responsible for geodesic
motion, \(h^{(A)}_{[\mu\nu]}\) accounts for the non-Riemannian twisting of the manifold induced
by the axial spin current \(J_A^\mu\).

The representation in Eq.~\eqref{eq:symm_component} is obtained by assuming that the
integration over the derivative index \(\lambda\) acts on the corresponding effective source term
\(S^{\rm eff}_{\mu\nu\lambda}\). This is physically justified under the assumption that the Green
function \(G(x,x')\) is slowly varying compared to the localized coherent spin distribution,
allowing it to be treated as approximately constant along the line integral defined
in~\eqref{eq:hmunu_reconstruct}.

\subsection{Yukawa--Screened Solutions}

The dynamical equations governing both the symmetric and antisymmetric sectors of the
spin--induced perturbation are of massive Klein--Gordon type. As a consequence, the
corresponding solutions are intrinsically short--ranged and exhibit Yukawa--type suppression.
This statement applies to individual microscopic contributions and does not refer to the
effective geometric structures that may emerge collectively at large distance scales. From
the definition of the reconstructed field
\begin{equation}
h_{(\mu\nu)}(x) = \int d^4x'\,G(x,x')\,J_{(\mu\nu)}(x'),
\end{equation}
and using the defining property of the Green function
\begin{equation}
(\Box - M^2)\,G(x,x') = \delta^{(4)}(x-x'),
\end{equation}
it follows directly that
\begin{equation}
(\Box - M^2)\,h_{(\mu\nu)}(x) = J_{(\mu\nu)}(x),
\end{equation}
where $M^2 = 8m^2$ and $J_{(\mu\nu)}$ denotes the symmetric projection of the effective coherent source defined in the previous section. In the static limit, the corresponding Green function
solution takes the standard Yukawa form,
\begin{equation}
h_{(\mu\nu)}(x) = \int d^3x'\,\frac{e^{-M|x-x'|}}{4\pi|x-x'|}\,J_{(\mu\nu)}(x').
\end{equation}
This expression shows that the metric perturbation induced by spin currents is exponentially
suppressed beyond the characteristic length scale $M^{-1}$. The gravitational response
associated with the symmetric sector is therefore intrinsically confined to short distances.

An analogous conclusion holds for the antisymmetric component $h_{[\mu\nu]}$, which obeys a
massive field equation sourced by the antisymmetric projection of the spinorial current. Its
solutions are likewise Yukawa--screened and remain localized near the spin sources.

As a result, both the direct axial torsional contribution and the effective metric
contribution associated with coherent axial-current domains are short--ranged at the level
of individual sources. Nevertheless, while each contribution is intrinsically confined, the
possibility that collective or organized configurations of spin--induced perturbations may
give rise to effective macroscopic imprints cannot be excluded and will be investigated in
future studies.

\subsection{Local Lorentz Invariance and Torsional Mass Scale}

The torsion--induced sector admits a nontrivial vacuum structure. As follows from the
Lorentz--invariant torsional Lagrangian~\cite{varani2025yukawa}, the Yukawa--type interaction
between chiral spinor currents and the torsional field allows for vacuum configurations
characterized by a nonvanishing expectation value,
\begin{equation}
\langle h_{\mu c|b}\{\gamma^\mu,\sigma^{cb}\}\rangle = v \neq 0.
\end{equation}
The presence of such a vacuum expectation value does not imply an explicit violation of
local Lorentz invariance at the level of the field equations, which remain fully Lorentz
covariant. Rather, it indicates that the spin--coupled geometric sector admits a nontrivial
vacuum configuration, characterized by a dynamically generated mass scale.

At the microscopic level, the dynamically generated mass implies that fluctuations of the
spin--induced field are massive and therefore Yukawa--suppressed. As a consequence, individual
torsional contributions are short--ranged and do not propagate as long--range dynamical modes.
In this regime, torsional degrees of freedom remain confined to local, non--Riemannian affine
effects.

The dynamically generated mass scale,
\begin{equation}
M_\mathrm{tors} \sim \sqrt{-\mu^2},
\end{equation}
thus controls the range of spin-induced geometrical effects. Here $\mu^2 < 0$ corresponds to
the symmetry-breaking parameter of the torsional potential, consistent with the spontaneous
symmetry breaking mechanism discussed in~\cite{varani2025yukawa}, which dynamically generates
the mass scale $M_\mathrm{tors}$. At energies well below $M_\mathrm{tors}$, individual
torsional perturbations are indeed suppressed. However, this short-range confinement does not
preclude the emergence of collective modes or topological configurations. In environments with
high spin density, the coherent superposition of local geometric responses can lead to
macroscopic effective fields. Simultaneously, the theory may admit non-trivial solutions, such
as torsional defects or global vacuum windings. In such cases, stability is guaranteed by the
global symmetry of the torsional sector. This provides a dual channel for spin-induced
physics: a massive, screened local response and a robust, potentially long-range topological
influence on the spacetime manifold.

\section{Results}

\subsection{General Spinorial Structure of the Source}
\label{sec:spinorial_structure}

While the symmetry properties of the reconstructed field were qualitatively introduced in
Section~\ref{sec:spincurrent}, we now revisit the algebraic structure of the source in
greater detail. This recall is essential to proceed with the comparison of those specific
cases where only the torsional degrees of freedom survive, as well as to provide a rigorous
foundation for the subsequent comparison with Weyl and Majorana spinors induced geometry.
We analyze the algebraic structure of the torsion--induced source term starting from
\begin{equation}
S_{\mu\nu\lambda} \propto 2i\,\bar\psi\,\{\gamma_\mu,\sigma_{\nu\lambda}\}\,\psi.
\label{eq:source_general}
\end{equation}

The four-dimensional Clifford identity for the anticommutator is
\begin{equation}
\{\gamma_\mu,\sigma_{\nu\lambda}\} = C_\epsilon\,\varepsilon_{\mu\nu\lambda\rho}\,
\gamma^\rho\gamma^5,
\label{eq:anticomm_correct}
\end{equation}
so that the source takes the purely axial form, up to an overall constant,
\begin{equation}
S_{\mu\nu\lambda} = \kappa_A\,\varepsilon_{\mu\nu\lambda\rho}\,J_A^\rho,
\qquad J_A^\rho = \bar\psi\,\gamma^\rho\gamma^5\,\psi.
\label{eq:source_axial2}
\end{equation}

The Levi--Civita structure implies
\begin{equation}
S_{\nu\mu\lambda} = \kappa_A\,\varepsilon_{\nu\mu\lambda\rho}\,J_A^\rho
= -\kappa_A\,\varepsilon_{\mu\nu\lambda\rho}\,J_A^\rho = -S_{\mu\nu\lambda},
\end{equation}
so the axial source is antisymmetric under $\mu\leftrightarrow\nu$.
For the purely axial microscopic source, the direct reconstructed contribution is therefore
antisymmetric in the first two indices and belongs to the torsional sector. A rank-two tensor
with an effective symmetric component arises only after including the coherent axial-current
response $S^{\rm eff}_{\mu\nu\lambda}$~\eqref{eq:Seff}.

Consequently, the bare microscopic Clifford source does not generate the
symmetric metric sector. The latter arises only from the coherent effective
response \(S^{\rm eff}_{\mu\nu\lambda}\) introduced in Eq.~\eqref{eq:Seff}.

If we restrict attention to the axial projection, the spinorial source term is
\begin{equation}
S_{\mu\nu\lambda} \propto \varepsilon_{\mu\nu\lambda\rho}\,J_A^\rho,
\label{eq:axial_source}
\end{equation}
where the axial current is defined as
\begin{equation}
J_A^\rho \equiv \bar\psi\,\gamma^\rho\gamma^5\,\psi.
\label{eq:axial_current}
\end{equation}
This structure is governed by the spin density, which corresponds to the pseudovector
representation of the Clifford algebra.

\subsection{Weyl Spinor Regime}
\label{sec:weyl}

In the Weyl regime, the spinor field satisfies the chirality eigenvalue equation
\begin{equation}
\gamma^5\,\psi = \pm\psi,
\end{equation}
which selects a definite chiral representation of the Lorentz group. In the four-component
formalism, a Dirac spinor can be decomposed into left- and right-handed components,
\begin{equation}
\psi = \begin{pmatrix}\psi_L\\\psi_R\end{pmatrix}.
\end{equation}
In this regime, spinorial currents are not independent. Instead, they are correlated through
chiral constraints. In particular, the vector and axial currents are defined as
\begin{equation}
J^\mu = \bar\psi\,\gamma^\mu\,\psi,\qquad
J_A^\mu = \bar\psi\,\gamma^\mu\gamma^5\,\psi.
\end{equation}
For massless chiral eigenstates, the currents satisfy
\begin{equation}
J_A^\mu = \pm J^\mu,
\label{eq:weyl_chiral_id}
\end{equation}
where the sign is determined by the chirality of the spinor
\((\gamma^5\psi=\pm\psi)\).

Starting from the microscopic axial spinorial source term
\begin{equation}
S^{(A)}_{\mu\nu\lambda}
\propto
\bar\psi\,\{\gamma_\mu,\sigma_{\nu\lambda}\}\,\psi,
\end{equation}
and using the Clifford algebra identity~\eqref{eq:anticomm_correct}, the spinorial
source in the Weyl regime becomes
\begin{equation}
S^{(A)}_{\mu\nu\lambda}
\propto
\pm\,\varepsilon_{\mu\nu\lambda\rho}\,J^\rho,
\label{eq:weyl_source}
\end{equation}
since \(J_A^\mu=\pm J^\mu\) in this limit. Therefore the axial anticommutator
source is controlled by the same chiral current that appears in the vector description.
The metric-like response, when present, is not generated directly by the bare microscopic
Clifford source, but is interpreted as an effective coherent response of this chiral current.

If the Weyl currents form a coherent domain, then
\begin{equation}
U^\mu
=
\langle J_A^\mu\rangle_{\rm coh}
=
\pm\langle J^\mu\rangle_{\rm coh}.
\end{equation}
The effective metric source becomes
\begin{equation}
S^{\rm eff}_{\mu\nu\lambda}
=
\pm\chi
\bigl(
g_{\mu\nu}\langle J_\lambda\rangle_{\rm coh}
-
g_{\mu\lambda}\langle J_\nu\rangle_{\rm coh}
\bigr).
\end{equation}
Therefore the effective metric structure in the Weyl regime is recovered as part of the
coherent effective response.

This regime is characterized by a chiral geometric configuration in which the torsional
sector is fundamentally controlled by the particle's chirality. In this massless limit, the
distinction between chirality and helicity vanishes, allowing the helicity of the physical
state to determine the orientation of the resulting torsional structure.

Furthermore, the Weyl regime is characterized by null-like propagation of fermionic currents,
\begin{equation}
J^\mu J_\mu = 0,
\end{equation}
indicating lightlike transport of spinorial degrees of freedom.

\subsection{From Weyl to Majorana Spinors}
\label{sec:majorana}

If we now consider a Majorana fermion, the reality condition
\begin{equation}
\psi = \psi^C = C\bar\psi^T
\end{equation}
imposes strong algebraic constraints on spinorial bilinears. The charge conjugation matrix
satisfies
\begin{equation}
C^{-1}\gamma^\mu C = -(\gamma^\mu)^T .
\end{equation}
The vector current can be written as
\begin{equation}
J^\mu
=
\bar\psi\,\gamma^\mu\,\psi
=
\psi^T C^{-1}\gamma^\mu\psi .
\end{equation}
Taking the transpose of the expression and using the charge conjugation property of the
gamma matrices yields
\begin{equation}
J^\mu = -J^\mu,
\end{equation}
which implies
\begin{equation}
J^\mu = 0 .
\end{equation}

The axial current
\begin{equation}
J_A^\mu
=
\bar\psi\,\gamma^\mu\gamma^5\,\psi
\end{equation}
does not vanish in general because
\begin{equation}
(\gamma^\mu\gamma^5)^T
=
+C^{-1}\gamma^\mu\gamma^5 C .
\end{equation}
The antisymmetric sector of the reconstructed field is thus sourced by the axial current,
\begin{equation}
h^{(A)}_{[\mu\nu]}(x)
\sim
\int dx'^\lambda
\int d^4x'\,G(x,x')\,
\varepsilon_{\mu\nu\lambda\rho}\,
J_A^\rho(x') .
\end{equation}

This confirms that the Majorana limit is the purely axial-torsional limit of the theory at
the microscopic level.
In both the Weyl and Majorana regimes, the Clifford anticommutator does not
generate a direct symmetric metric-like source, but selects the axial torsional
channel. A symmetric metric response may arise only at the coherent effective
level through the trace-vector response induced by the axial order parameter. In the Weyl case, this axial order parameter is locked to the chiral vector
current, \(U^\mu=\pm\langle J^\mu\rangle_{\rm coh}\). In the Majorana case,
instead, the microscopic vector current vanishes, \(J^\mu=0\), and \(U^\mu\)
is a purely axial collective order parameter.
If no macroscopic axial coherence is present, the Majorana regime remains
purely axial-torsional at the microscopic level, with no effective symmetric
metric response. This differs from the Weyl case in that the surviving axial
source is not locked to a nonvanishing chiral vector current, since the
Majorana vector current vanishes identically.
If a coherent Majorana domain is formed, however, one may define
\begin{equation}
U^\mu
=
\left\langle J_A^\mu\right\rangle_{\rm coh}.
\end{equation}
In this case, an effective metric response may arise at the collective level through the
trace-vector response
\begin{equation}
S^{\rm eff}_{\mu\nu\lambda}
=
\chi
\left(
g_{\mu\nu}U_\lambda
-
g_{\mu\lambda}U_\nu
\right).
\end{equation}
This does not contradict the vanishing of the microscopic vector current \(J^\mu=0\), since
the effective vector \(U^\mu\) is the coherent axial order parameter, not the Dirac vector
current.

The mechanism generating the effective metric sector is therefore formally the same in
the Weyl and Majorana regimes: in both cases it arises from the coherent trace-vector
response induced by the axial order parameter
\begin{equation}
U^\mu
=
\left\langle J_A^\mu\right\rangle_{\rm coh}.
\end{equation}
The difference lies in the nature of this order parameter. In the Weyl regime, one has
\begin{equation}
U^\mu
=
\left\langle J_A^\mu\right\rangle_{\rm coh}
=
\pm
\left\langle J^\mu\right\rangle_{\rm coh},
\end{equation}
so that the effective metric response is associated with a coherent null chiral current. In
the Majorana regime, instead, the microscopic vector current vanishes,
\begin{equation}
J^\mu=0,
\end{equation}
and \(U^\mu\) is a purely axial collective order parameter. Thus, if a metric response is
present in the Majorana regime, it is collective and axial in origin rather than driven by a
microscopic vector current.

Majorana fermions therefore represent a purely axial geometric regime at the microscopic
level: while the vector current vanishes, the torsional degrees of freedom are sourced
exclusively by the axial spin current. The resulting geometry is driven by intrinsic
spin-density correlations rather than by charge transport. In the presence of coherent axial
ordering, however, this purely torsional microscopic regime may become metrically active at
the collective level through an effective trace-vector response.

\subsection{Spin--Induced Corrections to Particle Motion}
\label{sec:motion}

We now analyze the implications of spin--induced perturbations for the motion of test
particles in flat spacetime. As established in the previous sections, the reconstructed
tensor \(h_{\mu\nu}\) admits a decomposition into symmetric and antisymmetric components,
\begin{equation}
h_{\mu\nu} = h_{(\mu\nu)} + h_{[\mu\nu]} .
\end{equation}
The symmetric component defines an effective metric sector when a coherent trace-vector
response is present, whereas the antisymmetric component encodes the torsional sector
directly sourced by the axial Clifford channel. The physical consequences depend on which
sector is dynamically realized.

\paragraph{Metric Response.}
When a coherent axial-current response generates a nonvanishing symmetric sector, test
particles follow metric geodesics determined by the effective spacetime metric
\begin{equation}
g^\mathrm{eff}_{\mu\nu}
=
\eta_{\mu\nu}
+
h^{\rm eff}_{(\mu\nu)} .
\end{equation}
In the present framework, this metric response is not generated directly by the bare
microscopic Clifford source. Rather, it arises from the coherent trace-vector response
associated with the axial order parameter \(U^\mu=\langle J_A^\mu\rangle_{\rm coh}\).

The motion of spinless particles in this effective metric background is therefore governed by
\begin{equation}
\frac{d^2x^\mu}{d\tau^2}
+
\left\{{}^\mu_{\alpha\beta}\right\}[g^\mathrm{eff}]
\dot x^\alpha\dot x^\beta
=
0,
\end{equation}
where the Levi--Civita connection is constructed from \(g^\mathrm{eff}_{\mu\nu}\). To first order,
\begin{equation}
\frac{d^2x^\mu}{d\tau^2}
=
-\delta\Gamma^\mu{}_{\alpha\beta}\,
\dot x^\alpha\dot x^\beta,
\end{equation}
with \(\delta\Gamma^\mu{}_{\alpha\beta}\) determined by derivatives of
\(h^{\rm eff}_{(\mu\nu)}\). Thus, whenever the coherent metric sector is present, the
spin-induced geometry produces a standard gravitational-like interaction consistent with
the weak equivalence principle at the metric coupling level.

\paragraph{Torsional Response.}
The antisymmetric sector \(h^{(A)}_{[\mu\nu]}\) is associated with the microscopic axial
Clifford source and therefore with the non-Riemannian torsional structure. The complete
connection may be written as
\begin{equation}
\Gamma^\mu{}_{\alpha\beta}
=
\left\{{}^\mu_{\alpha\beta}\right\}
+
K^\mu{}_{\alpha\beta},
\end{equation}
where \(\left\{{}^\mu_{\alpha\beta}\right\}\) is the Levi--Civita connection associated with
the symmetric metric sector, and \(K^\mu{}_{\alpha\beta}\) is the contorsion tensor derived
from the torsional sector. The contorsion tensor is related to the torsion tensor
\(T^\mu{}_{\alpha\beta}\) by
\begin{equation}
K_{\mu\alpha\beta}
=
\frac{1}{2}
\left(
T_{\mu\alpha\beta}
-
T_{\alpha\mu\beta}
-
T_{\beta\mu\alpha}
\right).
\end{equation}
In the linearized spin-induced regime, the torsion tensor may be represented schematically as
\begin{equation}
T_{\mu\alpha\beta}
\propto
\partial_\alpha h^{(A)}_{[\mu\beta]}
-
\partial_\beta h^{(A)}_{[\mu\alpha]} .
\end{equation}
Thus \(K^\mu{}_{\alpha\beta}\) encodes the affine non-Riemannian structure generated by the
spinorial source.

Trajectories associated with the full affine connection are described by the autoparallel
equation
\begin{equation}
\frac{d^2x^\mu}{d\tau^2}
+
\Gamma^\mu{}_{\alpha\beta}
\dot x^\alpha\dot x^\beta
=
0.
\end{equation}
Equivalently, this can be rewritten as a metric geodesic equation corrected by a torsional
term,
\begin{equation}
\frac{d^2x^\mu}{d\tau^2}
+
\left\{{}^\mu_{\alpha\beta}\right\}
\dot x^\alpha\dot x^\beta
=
-
K^\mu{}_{\alpha\beta}
\dot x^\alpha\dot x^\beta .
\end{equation}
If the torsion contains a trace-vector component, kinematical deviations from geodesic motion
may arise even for spin-blind probes. In the present construction, such a trace-vector sector
is associated with the coherent effective response, not with the bare microscopic axial Clifford
source.

Conversely, in the case of purely axial torsion, as realized microscopically in the Majorana
limit in the absence of coherent trace-vector ordering, the torsion tensor takes the form
\begin{equation}
T_{\mu\alpha\beta}
=
\varepsilon_{\mu\alpha\beta\rho}
A^\rho ,
\end{equation}
so that the contorsion tensor becomes totally antisymmetric,
\begin{equation}
K_{\mu\alpha\beta}
=
\frac{1}{2}
\varepsilon_{\mu\alpha\beta\rho}
A^\rho .
\end{equation}
Since \(K_{\mu\alpha\beta}=K_{[\mu\alpha\beta]}\), its contraction with the symmetric product
\(\dot x^\alpha\dot x^\beta\) vanishes identically,
\begin{equation}
K^\mu{}_{\alpha\beta}
\dot x^\alpha\dot x^\beta
=
0.
\end{equation}
Therefore purely axial torsion produces no deviation from geodesic motion for spinless test
particles, while retaining its coupling to intrinsic spin degrees of freedom.

This cancellation holds exclusively for spinless test particles. For spinning probes, the
relevant coupling is of the form
\begin{equation}
K_{\mu\alpha\beta}S^{\alpha\beta},
\end{equation}
where
\begin{equation}
S^{\alpha\beta}
=
\bar\psi\,\sigma^{\alpha\beta}\psi
\end{equation}
is the spin tensor of the probe. Since \(S^{\alpha\beta}\) is antisymmetric, the contraction
with a totally antisymmetric axial contorsion tensor does not vanish in general. Purely axial
torsion therefore remains dynamically active for polarized matter, producing spin-dependent
corrections to particle motion.

In this regime, torsion acts as a spin-selective geometric channel rather than as a universal
force field. It distinguishes the behavior of spinning matter from that of spin-blind test
bodies.

\paragraph{The Majorana Limit.}
This distinction becomes particularly clear in the Majorana configuration, where the Majorana
reality condition suppresses the microscopic vector current,
\begin{equation}
J^\mu=0,
\end{equation}
while the axial current \(J_A^\mu\) may remain nonvanishing. In the absence of macroscopic
axial coherence, no effective symmetric metric response is generated, and the leading-order
geometry remains purely axial-torsional. Spin-blind particles then propagate along
approximately inertial trajectories, whereas spinning probes may still experience
torsion-induced effects through direct spin--torsion coupling.

If, however, a coherent Majorana axial-current domain is formed, the order parameter
\begin{equation}
U^\mu
=
\left\langle J_A^\mu\right\rangle_{\rm coh}
\end{equation}
may induce an effective trace-vector response and hence a symmetric metric perturbation.
Thus, the Majorana regime is purely torsional at the microscopic level, but may become
metrically active at the collective level through coherent axial ordering.

The resulting picture is therefore twofold. In the general spinorial configuration, and in
coherent Weyl or Majorana domains, a symmetric metric sector may emerge and govern universal
geodesic motion. In the absence of such coherence, the direct microscopic Clifford source
generates only the axial torsional sector. The spinorial structure of the theory therefore
suggests a unified geometric picture in which metric and torsional properties emerge from
different levels of spinorial organization: microscopic axial spin sources generate torsion,
whereas coherent axial-current domains may induce an effective metric response.

\section{Conclusions}

In this work, we have developed a geometric framework in which spinor currents dynamically
source a generalized tensor field, inducing both torsional and effective metric perturbations
of the gravitational geometry~\cite{varani2025yukawa}. The fundamental dynamical variable is
formulated within the tetrad framework and projected onto spacetime indices. It obeys a
massive Klein--Gordon-type equation, leading to intrinsically short-ranged, Yukawa-suppressed
contributions.

A central structural result of the analysis is that the projection of the spin--induced field
onto spacetime indices yields a rank--two tensor without a definite symmetry under index
exchange. The symmetric and antisymmetric components therefore arise only after explicit
algebraic projection.

The antisymmetric sector is directly associated with axial torsional degrees of freedom and
non--Riemannian affine effects. The symmetric projection, when generated by a coherent
axial-current response, defines an effective metric perturbation governing geodesic motion.
Both the microscopic torsional response and the effective metric response inherit the massive,
Yukawa-suppressed character of the underlying spin-induced field.

The resulting particle dynamics exhibits a more nuanced separation of roles. Spinless test
particles are sensitive to the symmetric effective metric sector whenever it is present. They
remain fundamentally blind to the purely axial torsional sector due to the total antisymmetry
of the axial contorsion tensor. The direct microscopic source of the spin-induced geometry is
the axial current
\[
J_A^\rho = \bar\psi\,\gamma^\rho\gamma^5\,\psi,
\]
reflecting the Clifford algebra structure of the anticommutator
\(\{\gamma_\mu,\sigma_{\nu\lambda}\}\).

In the Majorana limit, the microscopic vector current vanishes and the direct Clifford source
is purely axial. In the absence of macroscopic axial coherence, the geometry is therefore
purely axial--torsional and no effective symmetric metric response is generated. The interaction
is thus shifted from a universal force-like dynamical effect to a topological and spin-dynamical
coupling. If coherent axial ordering is present, however, the Majorana regime may become
metrically active at the collective level through an effective trace-vector response.

Possible extensions include the investigation of non--linear regimes, topological torsional
configurations, and the conditions under which collective dynamical mechanisms may generate
macroscopic metric structures. In particular, the analysis of coherent spinorial behavior may
help to clarify how microscopic torsional excitations can give rise to macroscopic gravitational
effects, potentially bridging the gap between chiral neutrino dynamics and the large-scale
evolution of the universe~\cite{varani2025ricci}.

The model therefore has a two-level structure. At the microscopic level, the Clifford
anticommutator sources the axial torsional three-form
\begin{equation}
S^{(A)}_{\mu\nu\lambda}
=
\kappa_A\,\varepsilon_{\mu\nu\lambda\rho}\,J_A^\rho .
\end{equation}
At the collective level, coherent axial-current domains define an order vector
\begin{equation}
U^\mu
=
\left\langle J_A^\mu\right\rangle_{\rm coh},
\end{equation}
which induces the effective metric source
\begin{equation}
S^{\rm eff}_{\mu\nu\lambda}
=
\chi
\bigl(
g_{\mu\nu}U_\lambda
-
g_{\mu\lambda}U_\nu
\bigr).
\end{equation}
This separation preserves the emergence of both torsional and metric responses while keeping
distinct the microscopic axial channel and the collective effective metric response.
\appendix

\section*{Appendix A --- Spinorial Sources and Geometric Structure}

In the present framework, the fundamental geometric variables are formulated at the tetrad
level and may carry both symmetric and antisymmetric components when projected onto spacetime
indices. A consistent macroscopic description of spacetime geometry is obtained by projecting
the resulting rank--two field onto its symmetric sector, whenever an effective coherent metric
response is present. This symmetric sector defines the effective spacetime structure governing
gravitational observables, while the antisymmetric sector remains associated with torsional
and affine degrees of freedom.

In this Appendix, we collect a set of technical results underlying the construction of the
spin-induced geometry. These include the symmetrization procedure from tetrad perturbations,
the algebraic decomposition of the reconstructed rank--two field, and a brief discussion of the
emergence of topological structures associated with the non-Riemannian sector.

\subsection*{A.1 Symmetrization Procedure from Tetrad Perturbations}

In this Appendix we present the symmetrization procedure in a purely technical form, starting
from the tetrad formulation of the theory. The purpose is to show how a symmetric spacetime
metric is obtained from the symmetric projection of a general spin-induced perturbation.

The spacetime metric is defined in terms of the tetrad field as
\begin{equation}
g_{\mu\nu} = \eta_{ab}\,e^a{}_\mu\,e^b{}_\nu .
\end{equation}
In the linearized regime, the tetrad is expanded around the Minkowski background according to
\begin{equation}
e^a{}_\mu = \delta^a{}_\mu + \tfrac{1}{2}\,h^a{}_\mu ,
\end{equation}
where \(h^a{}_\mu\) denotes a general perturbation, carrying no a priori symmetry. Although the
spacetime metric is symmetric by definition, the tetrad perturbation \(h^a{}_\mu\) is a general
object and may contain antisymmetric components, which do not contribute to the metric but
encode independent affine and torsional degrees of freedom.

Substituting the tetrad expansion and retaining only first--order terms, one obtains
\begin{equation}
g_{\mu\nu}
=
\eta_{\mu\nu}
+
\tfrac{1}{2}
\left(
h_{\mu\nu}
+
h_{\nu\mu}
\right),
\end{equation}
where
\begin{equation}
h_{\mu\nu}
\equiv
\eta_{ab}\,\delta^a{}_\mu\,h^b{}_\nu .
\end{equation}
The metric can therefore be written as
\begin{equation}
g_{\mu\nu}
=
\eta_{\mu\nu}
+
h_{(\mu\nu)},
\qquad
h_{(\mu\nu)}
\equiv
\tfrac{1}{2}
\left(
h_{\mu\nu}
+
h_{\nu\mu}
\right).
\end{equation}
This shows that the symmetric projection of the perturbation is selected automatically by the
tetrad construction of the metric. The antisymmetric component \(h_{[\mu\nu]}\) drops out
identically from the effective metric sector and therefore does not contribute to the
macroscopic Riemannian geometry, while remaining dynamically relevant at the microscopic level
through torsional and affine degrees of freedom.

It is important to note that in standard General Relativity, the antisymmetric components of
the tetrad are typically associated with local Lorentz gauge degrees of freedom and can be
gauged away by a suitable choice of frame. However, in the present framework, the antisymmetric
sector is dynamically sourced by the axial spinorial channel and obeys a massive field equation.
Consequently, these components represent torsional degrees of freedom rather than a mere gauge
choice, encoding the non-Riemannian structure of the theory at the microscopic scale.

The non-symmetric structure of \(h_{\mu\nu}\) thus provides the degrees of freedom required to
describe both the effective metric response and the torsional sector. In particular, the
antisymmetric torsional sector may support nontrivial configurations, such as torsional defects,
vortices, or Skyrmion-like states, which would be absent in a purely symmetric Riemannian
description.

\subsection*{A.2 Algebraic Structure of the Axial Spinorial Source}

For completeness and self-consistency of the Appendix, we present the explicit derivation of
the algebraic structure of the spinorial source in the anticommutator channel.

The fundamental microscopic torsional source is defined as
\begin{equation}
S_{\mu\nu\lambda}
\propto
2i\,\bar\psi\,\{\gamma_\mu,\sigma_{\nu\lambda}\}\,\psi,
\qquad
\sigma_{\nu\lambda}
=
\tfrac{i}{2}[\gamma_\nu,\gamma_\lambda].
\end{equation}
Using the Clifford algebra identity for the anticommutator,
\begin{equation}
\{\gamma_\mu,\sigma_{\nu\lambda}\}
=
C_\epsilon\,\varepsilon_{\mu\nu\lambda\rho}\,
\gamma^\rho\gamma^5,
\label{eq:app_anticomm}
\end{equation}
the microscopic source takes the axial form, up to an overall constant,
\begin{equation}
S^{(A)}_{\mu\nu\lambda}
=
\kappa_A\,\varepsilon_{\mu\nu\lambda\rho}\,J_A^\rho,
\label{eq:app_axial_source}
\end{equation}
where
\begin{equation}
J_A^\rho
=
\bar\psi\,\gamma^\rho\gamma^5\,\psi
\end{equation}
is the axial current associated with the spinor field.

The source is antisymmetric in the last two indices,
\begin{equation}
S^{(A)}_{\mu\nu\lambda}
=
-
S^{(A)}_{\mu\lambda\nu},
\end{equation}
and also changes sign under \(\mu\leftrightarrow\nu\) due to the total antisymmetry of the
Levi--Civita tensor,
\begin{equation}
S^{(A)}_{\nu\mu\lambda}
=
-
S^{(A)}_{\mu\nu\lambda}.
\end{equation}
Therefore, the direct microscopic axial contribution belongs to the antisymmetric torsional
sector.

The effective rank--two source associated with the microscopic axial channel is defined by
integration over the derivative index,
\begin{equation}
J^{(A)}_{\mu\nu}(x)
=
\int dx'^\lambda\,S^{(A)}_{\mu\nu\lambda}(x').
\end{equation}
Substituting the explicit axial form gives
\begin{equation}
J^{(A)}_{\mu\nu}(x)
=
\kappa_A
\int dx'^\lambda\,
\varepsilon_{\mu\nu\lambda\rho}\,
J_A^\rho(x').
\end{equation}
Since
\begin{equation}
\varepsilon_{\nu\mu\lambda\rho}
=
-
\varepsilon_{\mu\nu\lambda\rho},
\end{equation}
the direct axial contribution is antisymmetric under \(\mu\leftrightarrow\nu\). Therefore,
by itself, it contributes to the antisymmetric torsional sector rather than to the symmetric
metric sector.

Recalling that
\begin{equation}
h^{(A)}_{\mu\nu|\lambda}(x)
=
\int d^4x'\,G(x,x')\,S^{(A)}_{\mu\nu\lambda}(x'),
\end{equation}
the corresponding reconstructed contribution is obtained by integration over the derivative
index,
\begin{equation}
h^{(A)}_{\mu\nu}(x)
=
\int dx'^\lambda\,h^{(A)}_{\mu\nu|\lambda}(x')
=
\int d^4x'\,G(x,x')\,J^{(A)}_{\mu\nu}(x').
\end{equation}
Its antisymmetric component is
\begin{equation}
h^{(A)}_{[\mu\nu]}
=
\tfrac{1}{2}
\left(
h^{(A)}_{\mu\nu}
-
h^{(A)}_{\nu\mu}
\right),
\end{equation}
and is directly associated with axial torsion through
\begin{equation}
S^{(A)}_{\mu\nu\lambda}
\propto
\varepsilon_{\mu\nu\lambda\rho}J_A^\rho.
\end{equation}

The symmetric metric sector is not generated by the bare microscopic axial Clifford source.
It is instead associated with the coherent axial-current response introduced in the main text,
\begin{equation}
S^{\rm eff}_{\mu\nu\lambda}
=
\chi
\left(
g_{\mu\nu}U_\lambda
-
g_{\mu\lambda}U_\nu
\right),
\qquad
U^\mu
=
\left\langle J_A^\mu\right\rangle_{\rm coh}.
\end{equation}
Thus, the microscopic anticommutator channel generates the axial torsional sector directly,
whereas the symmetric metric sector arises only as a collective coherent response.

\subsection*{A.3 Topological Structures from Axial Torsion}

We now consider the axial spin torsion. The breakdown of a strictly Riemannian structure
allows the emergence of nontrivial topological configurations. The antisymmetric sector of
the spin--induced perturbation, $h_{[\mu\nu]}$, encodes the axial spin structure of Majorana
fermions and maps internal spin degrees of freedom onto topological invariants of the
gravitational field. Specifically, this mapping is mediated by the purely axial nature of the
Majorana source: since the vector current vanishes identically in the Majorana regime, the
geometric sourcing is driven by the axial spin density $J_A^\mu$. The nonlinear dynamics of
the axial torsional sector admits stationary solutions with nontrivial topology, as shown in
Ref.~\cite{varani2025ricci}. As shown in Section~\ref{sec:motion}, purely axial torsion
couples exclusively to spinning matter through the contraction $K_{\mu\alpha\beta}S^{\alpha\beta}$,
providing the microscopic origin of the topological configurations discussed here.

Two classes of configurations are of particular relevance. Coherent domains of left--handed
and right--handed fermion currents support Skyrmion--like configurations characterized by
nonvanishing winding number, corresponding to mappings $\mathbb{R}^3\to S^3$~\cite{skyrme1961}.
Right--handed currents and left--right flip terms give rise to vortex--like solutions with
quantized phase winding, analogous to Abrikosov vortices~\cite{abrikosov1957}.

In the Majorana regime, these configurations carry a nontrivial topological charge that is
decoupled from the metric curvature, existing purely as a property of the torsional affine
structure.

\section*{Acknowledgments}

The author gratefully acknowledges Francisco Bulnes for useful discussions and encouragement.


\end{document}